# Interfacial-Water-Modulated Photoluminescence of Single-Layer WS$_2$ on Mica

Yanghee Kim[#1], Haneul Kang[#1], Myongin Song[1], Hyuksang Kwon[2] and Sunmin Ryu*[1]

[1]Department of Chemistry, Pohang University of Science and Technology (POSTECH), Pohang, Gyeongbuk 37673, Korea

*E-mail: sunryu@postech.ac.kr

## Abstract

Because of their bandgap tunability and strong light-matter interactions, two-dimensional (2D) semiconductors are considered promising candidates for next-generation optoelectronic devices. However, their photophysical properties are greatly affected by environments because of their 2D nature. In this work, we report that the photoluminescence (PL) of single-layer WS$_2$ is substantially affected by interfacial water that is inevitably present between itself and supporting mica substrates. Using PL spectroscopy and wide-field imaging, we show that the emission signals from A excitons and their negative trions decreased at distinctively different rates with increasing excitation power, which can be attributed to the more efficient annihilation between excitons than trions. By gas-controlled PL imaging, we also prove that interfacial water converts trions into excitons by depleting native negative charges through an oxygen reduction reaction, which renders excited WS$_2$ more susceptible to nonradiative decay via exciton-exciton annihilation. Understanding the roles of nanoscopic water in complex low-dimensional materials will eventually contribute to devising their novel functions and devices.

# 1. Introduction

Water confined in a reduced dimension exhibits unconventional properties because of the dominant molecular interactions with confining walls. The liquid−ice phase transition of water in carbon nanotubes is governed by the detailed balance between water-water and water-wall interactions.[1,2] Water flow through carbon nanotubes is a few orders of magnitude faster than what continuum hydrodynamics predicts.[3] Dielectric response of nm-thin water sandwiched between inorganic two-dimensional (2D) crystals is severely compromised because of wall-dictated structural ordering that suppresses molecular rotation.[4] Single (1L) and few-layer interfacial water sandwiched between graphene and substrates is rigid[5] and yet allows redox reactions that consume charge carriers of the graphene wall.[6,7] The ubiquitous presence of interfacial water within the assemblies of low-dimensional materials raises more questions beyond those answered already.

One question of priority and yet lacking clear understanding is how the electronic excitation in low-dimensional materials is affected by the tiny amount of water. The issue has been tackled by a few reports that used interfacial water entrapped between 2D semiconductors and solid substrates. Varghese et al.[8] observed that the photoluminescence (PL) of 1L $MoS_2$ on mica is modulated by the presence of interfacial water. The direction of charge transfer responsible for the PL change was dependent on the nature of substrates and the morphology of water layers. Park et al.[6] showed that sub-monolayer water accommodates the oxygen reduction reaction (ORR) that injects electrical holes in 1L $WS_2$ supported on $SiO_2$ and subsequently amplifies excitonic PL. By exploiting the PL characteristics of 1L $WS_2$, Kang et al.[9] observed the diffusion of molecular oxygen through the interfacial water layer in real-time. As demonstrated by these studies, the nanoscopic amount of water can have a substantial effect on the decay of excitons in low-dimensional materials. At the same time, it is also natural to expect that there can be other roles of interfacial water that have not been observed. Such exploration can be best performed using a TMD/mica system. Despite their low photoluminescence (PL) quantum yields, 1L transition metal dichalcogenides (TMDs) are a highly useful emitter because of the visible and NIR excitons with large binding energies.[10,11] Mica not only provides an atomically flat surface for 2D crystals[12] but also holds well-ordered water layers because of its crystallinity and large hydrophilicity.[13,14]

In this work, we report that interfacial water accelerates the relaxation of high-density excitons in 1L $WS_2$ supported on mica substrates. Sublinear power dependence of the PL signals originating from neutral and charged excitons showed that exciton-exciton annihilation (EEA) is an additional quenching channel and more dominating with interfacial water. Wide-field PL imaging revealed that water reduces net electron density in $WS_2$ by ORR. We conclude that subsequent charge-neutral $WS_2$ suffers more from EEA than charged $WS_2$ from EEA of charged excitons (trions). The findings of the current study will benefit fundamental research for excitonic dynamics in heterostructured systems and devising applications based on the control of excitonic fates.

## 2. Methods

**Preparation and treatments of samples.** Single-layer $WS_2$ samples were prepared at ambient conditions by mechanical exfoliation of bulk $WS_2$ crystals (2D semiconductors, Inc.) onto freshly cleaved mica substrates (Ted Pella, grade V1 muscovite mica).[14, 28] Fresh mica surfaces were also prepared by mechanical exfoliation. We identified the thickness of prepared samples by the optical contrast using an optical microscope (Nikon, LV100; 100× objective lens).[29] The relative humidity (RH) was measured to be in a range of 19% to 64% at room temperature of about 22 °C. Some $WS_2$ samples were deposited on mica maintained at 40 ~ 80 °C using a hot plate. The hot-transfer eased the release of the adhesive tape off the substrates and thus facilitated the exfoliation process. However, there was no meaningful correlation between the substrate's heating and the overall amount of interfacial water.

**Topographic measurements.** AFM (atomic force microscopy) characterizations were performed at ambient conditions in an amplitude-modulated non-contact mode.[14] Probe tips (MicroMasch, NSC-15) were oscillated with a nominal amplitude of 20 nm, and scan rates were between 0.3 and 0.5 Hz.

**Photoluminescence measurements.** PL was obtained by a homebuilt micro-PL spectrometer setup[6] with a grating of 300 grooves/mm at ambient conditions. A solid-state laser operating at a wavelength of 514 nm was used as the excitation source. The laser beam was focused onto a sample (FWHM of the focal spot ~ 0.60 µm) using a microscope objective (40X, numerical aperture = 0.60). The Back-scattered PL signal was collected with the same objective and guided to a spectrometer combined with a charge-coupled device. For wide-field PL imaging, the collimated laser beam was extended 3 times with a Galilean beam expander and then focused on the back-focal plane of the objective with a plano-convex lens (focal length = 500 mm).[6,9] For homogeneous excitation, the illumination area (FWHM ~ 45 µm) was maintained one order of magnitude larger than the typical sample. The PL signals in the range between 1.9 and 2.1 eV mostly contributed to the PL images.

## 3. Results and Discussion

Single layer (1L) WS$_2$ samples prepared on mica substrates typically spanned several microns across and gave a significant optical contrast, as shown in Fig. 1a. The AFM height images revealed plateau-like structures characteristic of interfacial water layers.[13,14] In Fig. 1b, most of the 1L WS$_2$ area was found to be ~0.7 nm elevated compared to the water-free areas (0WL) near the edges. Notably, the average elevation is twice the interplanar distance of the hexagonal ice, thus corresponding to water bilayers (2WL).[14] As confirmed for the graphene/mica system[13,14] the plateaus are attributed to water layers formed during the exfoliation step. The surface of mica is extremely hydrophilic and adsorbs water vapor in the air. As WS$_2$ layers are laminated on top of water-rich mica surfaces, the water is spread as flat interfacial layers to minimize mechanical deformation of 1L WS$_2$ with a large Young's modulus.[15] The sample in Fig. 1c showed 0 and 1WL regions in addition to clusters with a width of tens of nm. Even larger structures (blisters) were also found, as shown in Fig. 1d. Whereas the molecular-level structure of the interfacial layers is far from being understood, it is likely to be close to that of the hexagonal ice as depicted in Fig. 1e.[13,14,16]

The photoluminescence of 1L WS$_2$ is dominated by A exciton (X$^0$) and its charged species (X$^+$ or X$^-$).[6,17] Because of the prevalent native n-type doping,[17] negative trions are mostly observed from exfoliated samples.[6,9] In Fig. 2a (top), WS$_2$ directly supported on mica also contained the two features that are assembled into an apparently single asymmetric PL band at ~1.96 eV. The dissociation energy of $X^-$ turned out to be 18 to 33 meV when fitted with the sum of Lorentzian and Gaussian functions. The areal ratio between $X^-$ and $X^0$ (2.79:1) indicates the sample is doped with a significant density of electrons.[6] The polarity of the charges was verified by intentional chemical doping, as will be shown later. When the average power ($P_{exc}$) of the excitation beam was increased (middle and bottom of Fig. 2a), the overall PL intensity ($I_{tot}$) decreased notably: 55% decrease in $I_{tot}$ for 100 times increase in $P_{exc}$. We note that an even larger decrease was observed in the presence of interfacial water layers: ~85% for 1WL (Fig. 2b) and ~80% for 2WL (Fig. 2c). As will be discussed below, the reduction in the luminescence yield can be attributed to exciton-exciton annihilation (EEA)[18] that becomes more dominant at a higher density of excitons.[19-21] We also note that the line shape of the PL band changes with increasing $P_{exc}$. The fitted sub-components show that it is due to the differing power dependence of $X^-$ and $X^0$, as will be further discussed below.

In Fig. 3, we adopted a wide-field PL microscopy[6] to study the spatial variation of the power dependence of the excitonic emission. Because the two components were spectrally very close to each other, they were collected for imaging without spectral separation. The AFM image in Fig. 3a shows that the central 1L $WS_2$ region contains 1WL, whereas most of the edge areas are without interfacial water. Then, the PL images in Fig. 3a reveal that the 1WL-supported areas gave a stronger PL signal than the water-free region at the lowest $P_{exc}$ of 3.8 μW. With increasing $P_{exc}$ and simultaneous reduction of the corresponding exposure time, the signals from 1WL region decreased noticeably. Remarkably, the two regions showed an intensity inversion at the high $P_{exc}$. A similar observation was made for another sample with nWL (n = 0, 1 and 2) as shown in Fig. 3b. Compared to the 0WL areas near the edges, 1WL-supported $WS_2$ produced progressively less PL intensity with increasing $P_{exc}$.

Figure 3c presents the PL intensity selected from nWL-supported areas (Fig. 3b) as a function of $P_{exc}$ after normalization with respect to the fluence of excitation photons. The log-log intensity plot clearly shows the differing power dependence of nWL areas. Notably, 2WL areas also showed a high sensitivity towards $P_{exc}$, which was not readily visible in the PL images (Fig. 3b). In Fig. 3d, the same data set is given as a function of the photon fluence with each PL signal normalized to unity at the smallest fluence. The dotted line with a slope of unity represents an imaginary system with a constant PL quantum yield irrespective of $P_{exc}$. Then the fact that all the series in Fig. 3d have smaller slopes indicates that the PL process competes with fast nonradiative decay channels that are affected by the presence of interfacial water. As indicated by the PL images and Fig. 3d, single layers of water had the smallest slope of 0.60, which stands in stark contrast to the 0WL areas (slope of 0.81).

Notably, the presence of interfacial water enhances the nonradiative decay of the excitons in 1L $WS_2$. Among a few factors that will be discussed further below, we propose that the observed phenomenon relates to the hole doping of $WS_2$, which was induced by the oxygen reduction reaction (ORR) where water plays a key role.[6, 7] First of all, we verified that $WS_2$ samples supported on mica (e.g., Figs. 4a and 4b) are also hole-doped, and their excitonic emission is regulated by ORR like graphene and TMDs supported on hydrophilic silica substrates. Figures 4c and 4d present time-lapse PL and its enhancement images of 1L $WS_2$ obtained as a function of $O_2$-exposure time (t). The enhancement was defined as $(I_t - I_0)/I_0$, where $I_0$ corresponds to the PL intensity at t = 0. To control the gaseous environment, samples were mounted in a gas-flow optical cell. Because the cell was purged with high-purity Ar for two hours before the introduction of $O_2$, the PL enhancement can be attributed to the action by $O_2$. The earliest image (t = 15 s) in Fig. 4d indicates a slight enhancement near some of the edges, and a more prominent change occurred within 2 min. The following several minutes observed a clear edge-to-center propagation of the enhancement fronts. This feature indicates that the PL enhancement is caused by molecular species that diffuse through the $WS_2$/mica interface from the atmosphere. Considering its reversibility upon exposure to Ar (Fig. S1), the reaction behind the overall change is identical to the ORR that was observed for $WS_2$ supported on $SiO_2$: $O_2 + 4H^+ + 4e^- \rightleftharpoons 2H_2O$ (acidic) and $O_2 + 2H_2O + 4e^- \rightleftharpoons 4OH^-$ (basic).[7] Whereas the electrons are provided by $WS_2$, water required for the electrochemical reaction can be readily found at the interface formed by hydrophilic substrates like $SiO_2$ and mica in this study. Even the 0WL areas near the edges of Fig. 4b include numerous tiny water clusters, possibly with smaller clusters that are not seen in the AFM images.

When natively n-doped $WS_2$ is hole-doped by the ORR, the net charge density decreases, and the PL signals will be dominated by neutral excitons rather than trions.[6] Indeed, Fig. 2 shows that the contribution of $X^0$ is larger in the water-supported areas than in the water-free regions.

Then the nature of EEA readily explains the intensity inversion observed for 1WL-supported and water-free areas. In an EEA process, two excitons collide with each other, and one decays by dumping its energy into the other, which is simultaneously excited further and eventually relaxes nonradiatively.[18] Because of its bimolecularity, EEA becomes more contributing at a higher density of excitons, in other words, higher $P_{exc}$. Compared to EEA, trion-trion annihilation (TTA) is less efficient because of the repulsive interaction between the charged excitons as depicted in Fig. 3e.[22] Indeed, the PL spectra in Fig. 2 show that the intensity of $X^0$ decreases more than $X^-$ with increasing $P_{exc}$ irrespective of the presence of the interfacial water. Then, the stronger power-dependence of water-supported areas can be attributed to the fact that their PL band is dominated by $X^0$, not $X^-$.

We note other factors may be responsible for the observations. First, one may consider that the efficient dielectric screening of water is somehow related. Indeed, energetic relaxation by various solvents has been observed for excitons in TMDs.[23] However, it has recently been shown that the dielectric constant of water is greatly lowered because of water-wall interactions with decreasing dimension.[4] In addition, a few monolayers of water is highly ordered on the surface mica,[16, 24] which should lead to further reduction in dielectric screening. Second, interfacial water may induce various degree of structural corrugation in $WS_2$ because 2D materials are prone to out-of-plane deformation and tend to maximize van der Waals interactions with underlying substrates.[25, 26] Such deformed lattice sites may assist nonradiative decay by acting as centers for momentum exchange required for various scattering processes as shown in the double resonant Raman scattering of graphene.[27] We also note that the morphology of interfacial water may affect the exciton-trion equilibrium. As shown by the samples containing water clusters (Fig. 1c), non-flat and three-dimensional forms of interfacial water are more likely to induce severe deformation of $WS_2$ and subsequently larger interfacial voids, which serve as an efficient diffusion channel for $O_2$ required for ORR.[9] Compared to hexagonal ice that can form epitaxially on mica, less structured water is also more suitable for ORR as a solvent because of its enhanced fluidity.[16] Indeed, the PL and enhancement images in Fig. 4 show that $WS_2$ areas with high roughness accommodate more active ORR, which leads to a larger exciton/trion ratio and stronger PL signals.

## 4. Conclusions

In this work, we investigated the effects of interfacial water on the excitonic behaviors of 1L $WS_2$ supported on mica. Using mechanical exfoliation, $WS_2$ samples could be generated with well-defined mono and bilayers of interfacial water in addition to essentially water-free areas. PL spectroscopy and wide-field imaging showed that the emission signals from A excitons and their negative trions decreased at distinctively different rates with increasing excitation power. The more pronounced power dependence of the former is attributed to the larger annihilation probability between excitons than trions. Atmosphere-controlled PL imaging also revealed that interfacial water charge-neutralizes natively n-doped $WS_2$ via ORR and essentially converts trions into excitons, the latter of which decay nonradiatively more efficiently than the former. This study will shed light on the roles of nanoscopic water in complex low-dimensional materials and eventually contribute to devising their novel functions and devices.

ASSOCIATED CONTENT

Supporting information

Supplementary data associated with this article can be found, in the online version.

AUTHOR INFORMATION

Corresponding Authors

*E-mail: sunryu@postech.ac.kr

#These authors contributed equally.

Notes

The authors declare no conflict of interest.

## ACKNOWLEDGMENTS

This work was supported by the National Research Foundation of Korea (NRF-2022R1A4A1033247 and NRF-2021R1A6A1A10042944), Samsung Science and Technology Foundation under Project Number SSTF-BA1702-08 and Samsung Electronics Co., Ltd (IO201215-08191-01).

## REFERENCES

1.	Pascal, T.A., W.A. Goddard, and Y. Jung, Entropy and the driving force for the filling of carbon nanotubes with water. Proceedings of the National Academy of Sciences of the United States of America, 2011. 108(29): 11794-11798.

2.	Koga, K., G.T. Gao, H. Tanaka, and X.C. Zeng, Formation of ordered ice nanotubes inside carbon nanotubes. Nature, 2001. 412(6849): 802-805.

3.	Holt, J.K., H.G. Park, Y.M. Wang, M. Stadermann, A.B. Artyukhin, C.P. Grigoropoulos, A. Noy, and O. Bakajin, Fast mass transport through sub-2-nanometer carbon nanotubes. Science, 2006. 312(5776): 1034-1037.

4.	Fumagalli, L., A. Esfandiar, R. Fabregas, S. Hu, P. Ares, A. Janardanan, Q. Yang, B. Radha, T. Taniguchi, K. Watanabe, G. Gomila, K.S. Novoselov, and A.K. Geim, Anomalously low dielectric constant of confined water. Science, 2018. 360(6395): 1339-1342.

5.	Lee, D., G. Ahn, and S. Ryu, Two-dimensional water diffusion at a graphene-silica interface. J. Am. Chem. Soc., 2014. 136(18): 6634-6642.

6.	Park, K., H. Kang, S. Koo, D. Lee, and S. Ryu, Redox-governed charge doping dictated by interfacial diffusion in two-dimensional materials. Nat. Commun., 2019. 10(1): 4931.

7.	Kang, H., K. Park, and S. Ryu, Optical Imaging of Redox and Molecular Diffusion in 2D van der Waals Space. Accounts of Chemical Research, 2022. 55(1): 44-55.

8.	Varghese, J.O., P. Agbo, A.M. Sutherland, V.W. Brar, G.R. Rossman, H.B. Gray, and J.R. Heath, The Influence of Water on the Optical Properties of Single-Layer Molybdenum Disulfide. Advanced Materials, 2015. 27(17): 2734-2740.

9.	Kang, H. and S. Ryu, Optical Imaging of Chemically and Geometrically Controlled Interfacial Diffusion and Redox in 2D van der Waals Space. The Journal of Physical Chemistry C, 2021. 125(30): 16819-16826.

10.	Mak, K.F., C. Lee, J. Hone, J. Shan, and T.F. Heinz, Atomically thin $MoS_2$: A new direct-gap semiconductor. Physical Review Letters, 2010. 105(13): 136805.

11.	Splendiani, A., L. Sun, Y.B. Zhang, T.S. Li, J. Kim, C.Y. Chim, G. Galli, and F. Wang, Emerging photoluminescence in monolayer $MoS_2$. Nano Letters, 2010. 10(4): 1271-1275.

12. Lui, C.H., L. Liu, K.F. Mak, G.W. Flynn, and T.F. Heinz, Ultraflat graphene. Nature, 2009. 462: 339-341.

13. Xu, K., P.G. Cao, and J.R. Heath, Graphene Visualizes the First Water Adlayers on Mica at Ambient Conditions. Science, 2010. 329(5996): 1188-1191.

14. Shim, J., C.H. Lui, T.Y. Ko, Y.-J. Yu, P. Kim, T.F. Heinz, and S. Ryu, Water-Gated Charge Doping of Graphene Induced by Mica Substrates. Nano Letters, 2012. 12(2): 648-654.

15. Falin, A., M. Holwill, H. Lv, W. Gan, J. Cheng, R. Zhang, D. Qian, M.R. Barnett, E.J.G. Santos, K.S. Novoselov, T. Tao, X. Wu, and L.H. Li, Mechanical Properties of Atomically Thin Tungsten Dichalcogenides: WS2, WSe2, and WTe2. ACS Nano, 2021. 15(2): 2600-2610.

16. He, K.T., J.D. Wood, G.P. Doidge, E. Pop, and J.W. Lyding, Scanning Tunneling Microscopy Study and Nanomanipulation of Graphene-Coated Water on Mica. Nano Letters, 2012. 12(6): 2665-2672.

17. Mak, K.F., K. He, C. Lee, G.H. Lee, J. Hone, T.F. Heinz, and J. Shan, Tightly bound trions in monolayer MoS2. Nature Materials, 2013. 12(3): 207-211.

18. Uddin, S.Z., E. Rabani, and A. Javey, Universal Inverse Scaling of Exciton–Exciton Annihilation Coefficient with Exciton Lifetime. Nano Letters, 2021. 21(1): 424-429.

19. Yu, Y., Y. Yu, C. Xu, A. Barrette, K. Gundogdu, and L. Cao, Fundamental limits of exciton-exciton annihilation for light emission in transition metal dichalcogenide monolayers. Physical Review B, 2016. 93(20): 201111.

20. Hoshi, Y., T. Kuroda, M. Okada, R. Moriya, S. Masubuchi, K. Watanabe, T. Taniguchi, R. Kitaura, and T. Machida, Suppression of exciton-exciton annihilation in tungsten disulfide monolayers encapsulated by hexagonal boron nitrides. Physical Review B, 2017. 95(24): 241403.

21. Lee, Y., G. Ghimire, S. Roy, Y. Kim, C. Seo, A.K. Sood, J.I. Jang, and J. Kim, Impeding Exciton–Exciton Annihilation in Monolayer WS2 by Laser Irradiation. ACS Photonics, 2018. 5(7): 2904-2911.

22. Chatterjee, S., G. Gupta, S. Das, K. Watanabe, T. Taniguchi, and K. Majumdar, Trion-trion annihilation in monolayer WS2. Physical Review B, 2022. 105(12): L121409.

23. Lin, Y., X. Ling, L. Yu, S. Huang, A.L. Hsu, Y.-H. Lee, J. Kong, M.S. Dresselhaus, and T. Palacios, Dielectric Screening of Excitons and Trions in Single-Layer MoS2. Nano Letters, 2014. 14(10): 5569-5576.

24. Miranda, P.B., L. Xu, Y.R. Shen, and M. Salmeron, Icelike Water Monolayer Adsorbed on Mica at Room Temperature. Phys. Rev. Lett., 1998. 81: 5876-5879.

25. Stolyarova, E., D. Stolyarov, K. Bolotin, S. Ryu, L. Liu, K.T. Rim, M. Klima, M. Hybertsen, I. Pogorelsky, I. Pavlishin, K. Kusche, J. Hone, P. Kim, H.L. Stormer, V. Yakimenko, and G. Flynn, Observation of Graphene Bubbles and Effective Mass Transport under Graphene Films. Nano Letters, 2009. 9(1): 332-337.

26. Locatelli, A., K.R. Knox, D. Cvetko, T.O. Mentes, M.A. Nino, S.C. Wang, M.B. Yilmaz, P. Kim, R.M. Osgood, and A. Morgante, Corrugation in Exfoliated Graphene: An Electron Microscopy and Diffraction Study. ACS Nano, 2010. 4(8): 4879-4889.

27. Ferrari, A.C. and D.M. Basko, Raman spectroscopy as a versatile tool for studying the properties of graphene. Nature Nanotechnology, 2013. 8(4): 235-246.

28. Park, K. and S. Ryu, Dual-channel charge transfer doping of graphene by sulfuric acid. Bulletin of the Korean Chemical Society, 2022. 43(1): 40-43.

29. Lee, J., T.Y. Ko, J.H. Kim, H. Bark, B. Kang, S.G. Jung, T. Park, Z. Lee, S. Ryu, and C. Lee, Structural and optical properties of single-and few-layer magnetic semiconductor CrPS4. ACS Nano, 2017. 11(11): 10935-10944.

FIGURE CAPTIONS

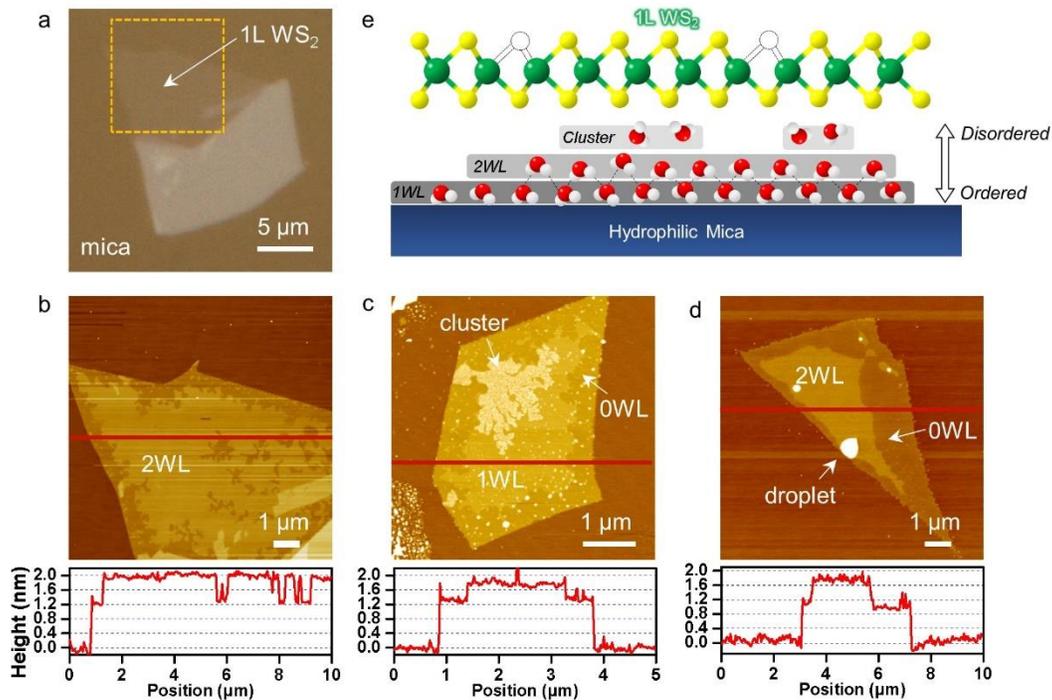

**Figure 1. Morphology of interfacial water layers.** (a & b) Optical micrograph (a) and AFM height image and profile (b) of 1L $WS_2$ on mica. The AFM image showing 2L interfacial water (2WL) was obtained from the orange square in (a). The height profile in (b) was taken along the red line in (b). (c & d) AFM images and height profiles of two additional samples containing various forms of interfacial water. (e) Scheme of 1L $WS_2$/water/mica.

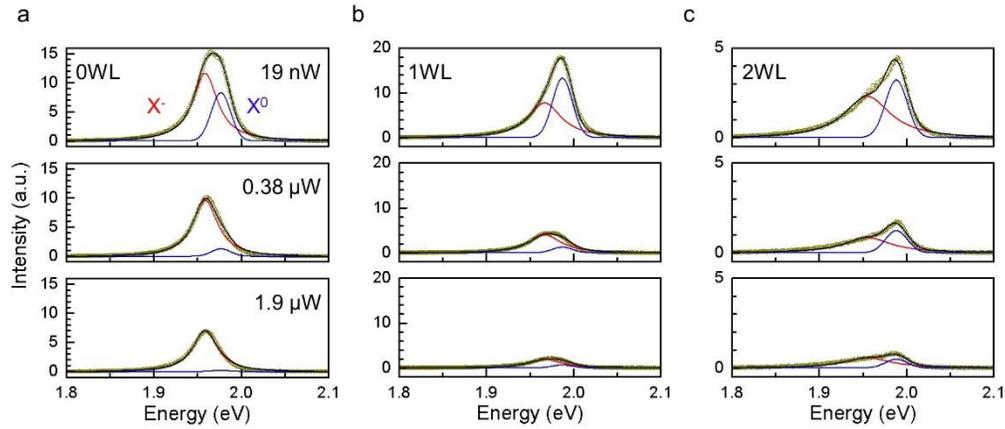

**Figure 2. Power-dependent PL spectra of WS$_2$/water/mica.** (a ~ c) A exciton PL spectra obtained from water-free (a), 1WL (b) and 2WL (c) areas. For each, the average power (P$_{exc}$) of excitation beam was set to 0.019, 0.38 and 1.9 μW (from top to bottom). Neutral exciton (X$^0$) and negative trion (X$^-$) are fitted with a Gaussian (blue) and a Lorentzian (red) functions, respectively.

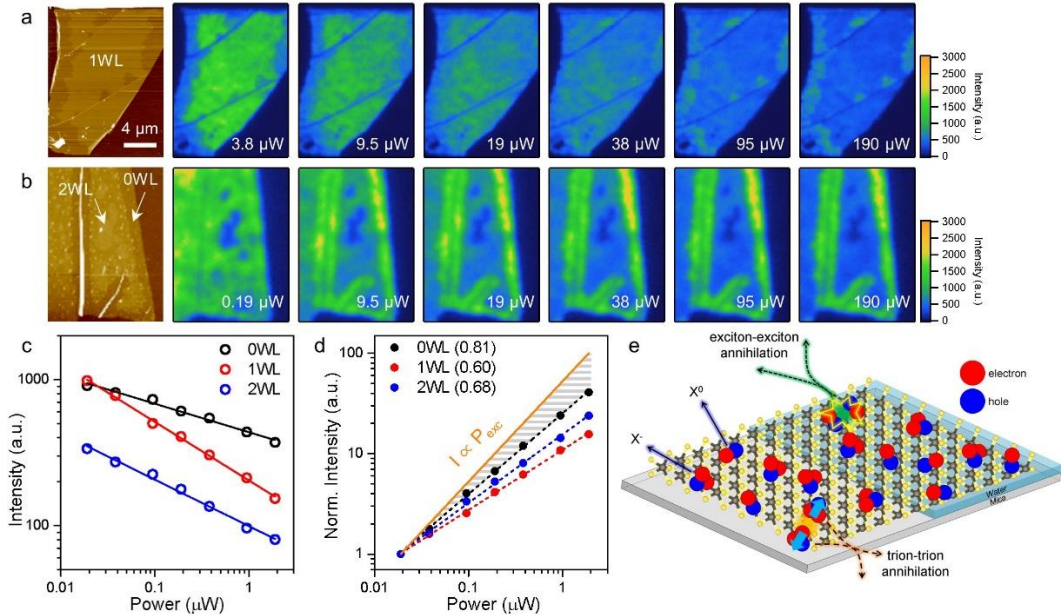

**Figure 3. Wide-field PL imaging of power-dependent intensity inversion.** (a) AFM (leftmost) and PL (the others) images of WS$_2$/mica with 0WL and 1WL areas. (b) AFM (leftmost) and PL (the others) images of WS$_2$/mica with 0WL, 1WL and 2WL areas. Whereas P$_{exc}$ was increased in (a) and (b), the corresponding exposure time was decreased reciprocally for a constant photon

fluence. (c) Integrated PL intensity of $X^-$ and $X^0$ peaks for 0WL, 1WL and 2WL areas of (b). Each PL spectrum was obtained for a constant photon fluence while varying $P_{exc}$. Note that $P_{exc}$ for the PL imaging was two orders of magnitude larger than that for the PL spectra in (c) because of a large illumination area for the imaging. (d) Normalized PL intensity (I) from (c) given as a function of $P_{exc}$. Each data point in (c) was divided by its exposure time and then normalized with respect to the value at the lowest $P_{exc}$. (e) Schematic representation of exciton-exciton annihilation (EEA) and trion-trion annihilation (TTA) processes in $WS_2$/water/mica.

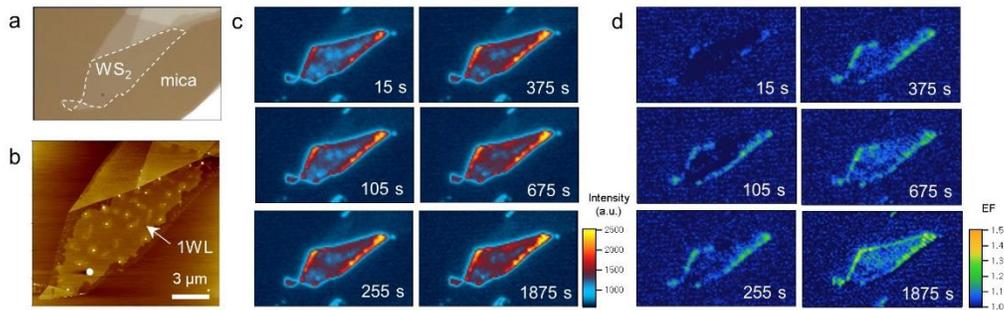

**Figure 4.** (a & b) Optical micrograph (a) and AFM height image (b) of 1L $WS_2$/mica. (c & d) Time-lapse PL images (c) and PL enhancement images (d) of the sample in (a). The sample was pre-equilibrated with Ar gas for 2 h before exposure to Ar:$O_2$ mixed gas at time zero. The enhancement factor (EF) was calculated by normalizing each PL image with respect to that at time zero.